# Quantum Sensing of Insulator-to-Metal Transitions in a Mott Insulator

Nathan J. McLaughlin[1, †], Yoav Kalcheim[1, †], Albert Suceava[1], Hailong Wang[2], Ivan K. Schuller[1,2], Chunhui Rita Du[1,2]

[1]Department of Physics, University of California San Diego, La Jolla, California 92093
[2]Center for Memory and Recording Research, University of California, San Diego, La Jolla, California 92093

[†]These authors made equal contributions to this work

Nitrogen vacancy (NV) centers, optically-active atomic defects in diamond, have attracted tremendous interest for quantum sensing, network, and computing applications due to their excellent quantum coherence and remarkable versatility in a real, ambient environment. Taking advantage of these strengths, we report on NV-based local sensing of the electrically driven insulator-to-metal transition (IMT) in a proximal Mott insulator. We studied the resistive switching properties of both pristine and ion-irradiated $VO_2$ thin film devices by performing optically detected NV electron spin resonance measurements. These measurements probe the *local* temperature and magnetic field in electrically biased $VO_2$ devices, which are in agreement with the *global* transport measurement results. In pristine devices, the electrically-driven IMT proceeds through Joule heating up to the transition temperature while in ion-irradiated devices, the transition occurs non-thermally, well below the transition temperature. Our results provide the first direct evidence for non-thermal electrically induced IMT in a Mott insulator, highlighting the significant opportunities offered by NV quantum sensors in exploring nanoscale thermal and electrical behaviors in Mott materials.

As technology begins to reach the fundamental limitations of the von Neumann computing architectures, new paradigms are urgently required to improve the processing speed, data storage capacity, and energy efficiency for next-generation information technologies.[1] Over the past decade, new approaches to information processing, such as quantum,[2,3] neuromorphic,[4–6] and material implication,[7] have been very active fields of research. Among them, neuromorphic computing is designed to mimic the highly interconnected structure of biological neural systems like the human brain.[8] Computations are spread out across an array of artificial neurons, synapses, and dendrites to provide vast improvements in global pattern recognition, complex input handling, and artificial intelligence capabilities.[8–10] Insulator-to-metal transitions (IMTs) in Mott materials featuring first-order, threshold firing type resistive switching behavior are naturally relevant in this context due to their significant potential for implementing artificial spiking neurons in neuromorphic circuits.[11–15] Thus, great efforts have been devoted to understanding and controlling the mechanism for resistive switching in materials exhibiting an IMT.[16–20] Successful applications of IMTs to emergent neuromorphic computing platforms require advances in theory, material discovery, and equally importantly, a detailed knowledge of the local electrical and thermal properties of Mott materials down to the nanoscale regime.

Conventional research on IMTs has been mainly focused on *global* electrical transport measurements as well as structural characterizations, rendering limited information on the local properties of the studied materials. To address this challenge, here, we utilized nitrogen vacancy (NV) centers, optically-active atomic defects in diamond that act as single-spin sensors,[21] to perform local quantum sensing of the voltage-induced IMT in a prototypical Mott material: vanadium dioxide $VO_2$.[5,12,22] Notably, the measured magnetic fields generated by the $VO_2$ devices exhibit a characteristic step-like jump around a "critical" electric current, in agreement with the formation of conducting filaments during the voltage-induced IMT.[5,23] The temperature profile we observed for both pristine and ion-irradiated $VO_2$ films are explained by thermal[24] and non-thermal origins.[25] We expect that the presented NV-based quantum sensing platform can be extended naturally to other Mott insulators, offering a new perspective to reveal the *local* thermal and electrical behaviors in Mott-material-based neuromorphic devices.

The measurement platform and device structure used are illustrated in Fig. 1(a). We have grown 170-nm-thick $VO_2$ films on $Al_2O_3$ (012) substrates by radio-frequency magnetron sputtering.[5] Two 125-nm-thick Ti/Au electrodes were fabricated with a separation of 10 μm on top of a $VO_2$ film for electrical transport measurements. Patterned diamond nanobeams containing individually addressable NV centers[26] were transferred on top of the $VO_2$ film and positioned between the two Au electrical contacts as shown by a scanning electron microscope (SEM) image in Fig. 1(b). The diamond nanobeam has the shape of an equilateral triangular prism with dimensions of 500 nm × 500 nm × 10 μm. The NV-to-sample distance was estimated to be ~100 nm,[27] ensuring sufficient thermal and field sensitivity. An Au stripline was fabricated next to the electrical contacts to provide microwave control of the NV spin states.[27,28] The whole sample was mounted on a heating stage to allow for a precise adjustment of the base temperature. IMTs can be thermally and/or electrically triggered in the $VO_2$ device, accompanied by orders of magnitude decrease in electrical resistivity.[24,25,29] The NV center positioned on top of the $VO_2$ film serves as a local probe of the temperature and magnetic field at the NV site. Taking advantage of their excellent quantum coherence and single-spin sensitivity,[21,30,31] NV centers have been demonstrated to be a transformative tool in exploring the local magnetic, electric and thermal features of a variety of materials with unprecedented sensitivity and spatial resolution.[30,32–35] Figure 1(c) shows a confocal optical image of an NV center positioned on the symmetry axis of the two Au electrical contacts, demonstrating the single-spin addressability of our measurement system.

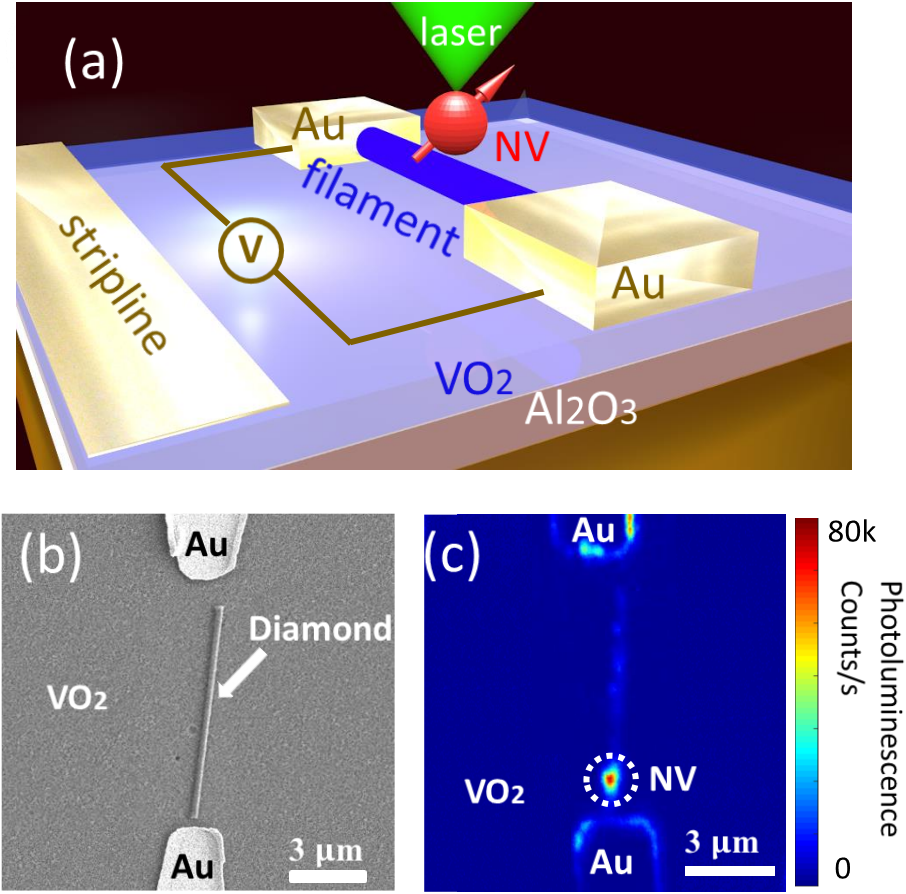

**Figure 1.** (a) Schematic of a prepared NV-VO$_2$ device consisting of two Au electrodes and an Au microwave stripline fabricated on top of a 170-nm-thick VO$_2$ thin film. A diamond nanobeam containing individually addressable NV centers is transferred on top of the VO$_2$ film to perform local thermal and field sensing of IMT. (b) A SEM image showing a patterned diamond nanobeam situated between the Au electrical contacts. (c) A photoluminescence image showing a diamond nanobeam containing a single NV spin positioned between the two the Au electrical contacts.

We first performed electrical transport measurements to characterize the temperature induced IMT in the VO$_2$ device. Figure 2(a) shows the resistance of the VO$_2$ device as a function of base temperature. The blue and the red curves correspond to the heating and cooling branches, respectively. The device features a characteristic IMT around 335 K, in agreement with the critical temperature $T_c$ reported in previous work.[5,36] After establishing the resistance-temperature profile, we demonstrate the electrically-induced IMT in the VO$_2$ device. In these measurements, the base temperature was maintained at a constant value below $T_c$, while an electric current $I_{dc}$ applied in the VO$_2$ channel was slowly varied and the voltage was simultaneously measured between the two Au contacts. At a certain critical current $I_c$, a conducting filament stretching between the tips of the two Au contacts was formed in the VO$_2$ film, leading to a sudden and significant drop of the measured voltages as shown in Fig. 2(b). This is due to a metallic filament forming between the Au contacts which is visible by optical microscopy due to the change in optical constants [Fig.

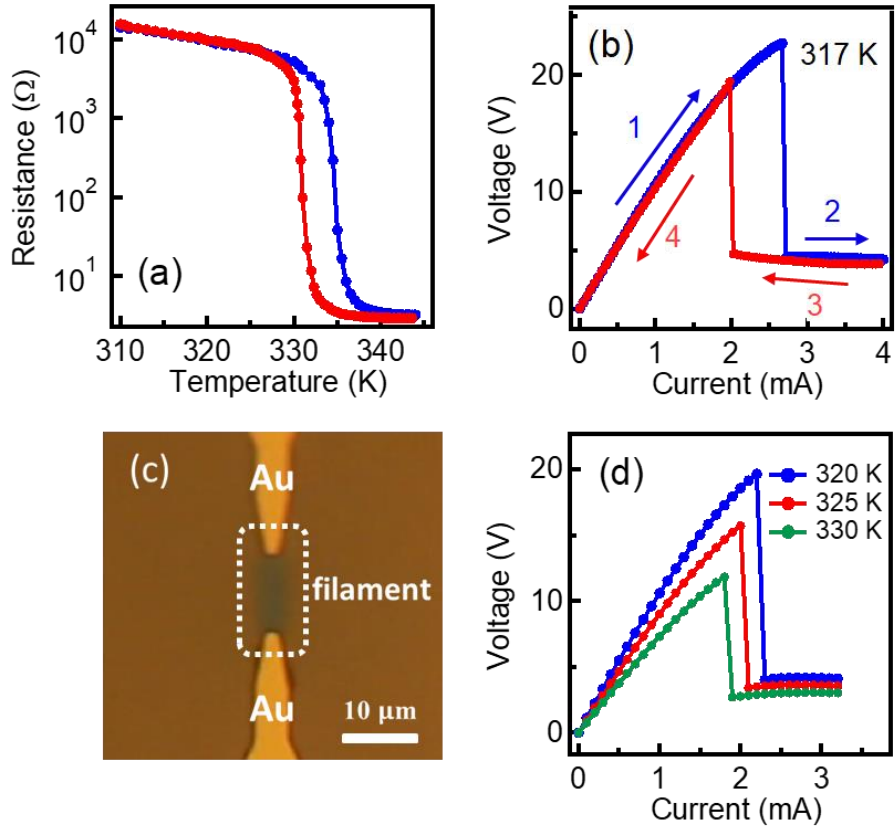

**Figure 2**. (a) Resistance of a pristine $VO_2$ device measured as a function the base temperature. Blue and red curves correspond to the heating and cooling branches, respectively. (b) The voltage (*V*) measured as a function of applied current ($I_{dc}$) at a base temperature of 317 K. Blue and red curves correspond to increasing and decreasing current, respectively. When sweeping the current down, the filament shrinks and supports the metallic state with a lower critical current. (c) An optical image showing the formation of a conducting filament (dark color) between the two Au contacts. (d) Electric voltage (*V*) measured as a function of the applied current ($I_{dc}$) at three different base temperatures.

2(c)]. Note that the magnitudes of the critical currents and voltages of the electrically-induced IMT depend on the base temperature of the device. The closer the base temperature is to $T_c$, the smaller the electric power is required to activate the resistive switching as illustrated in Fig. 2(d). When the base temperature of the device is too far below $T_c$, it is not possible to trigger voltage-induced IMTs in the $VO_2$ film without reaching excessively high voltages, where irreversible phase variations into other vanadium oxide compounds could occur.[29] In the following NV measurements, the base temperature of the sample is set to be above 295 K to avoid irreversible damage to the $VO_2$ devices.

    Next, we demonstrate the NV center's ability to accurately detect the local temperature and magnetic field environment of the $VO_2$ device during the voltage-induced IMTs. The negatively charged NV state has an *S* = 1 electron spin with a spin triplet ground state. Figure 3(a) shows the energy levels of an NV spin as a function of an external magnetic field ($B_{//}$) applied along the NV-axis. When $B_{//} = 0$, the $m_s = \pm 1$ states of the NV spin are degenerate with a characteristic NV zero-splitting frequency. At a certain external field ($B_{//} \neq 0$), Zeeman coupling separates the $m_s =$

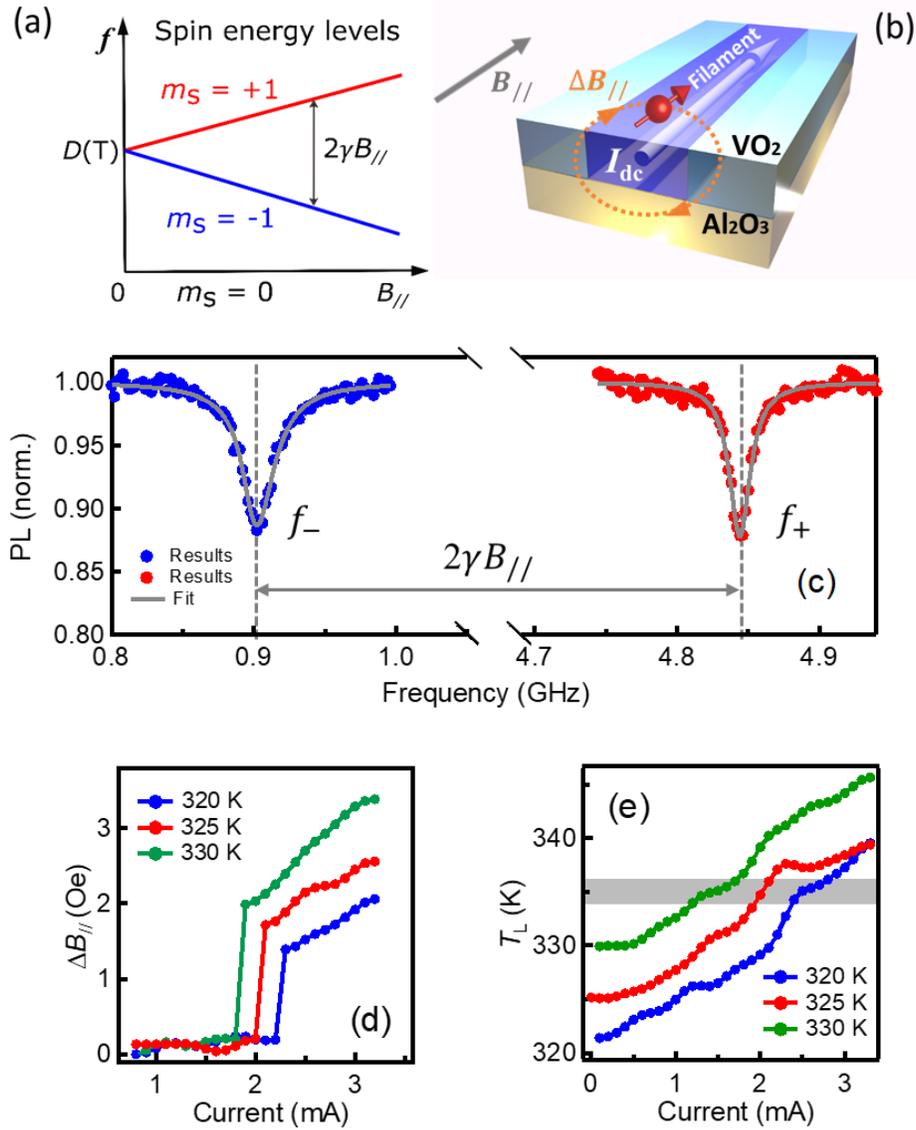

**Figure 3**. (a) An energy diagram of an NV spin as a function of an external magnetic field $B_{//}$ applied along the NV-axis. $D(T)$ represents the NV zero-splitting frequency, which varies as a function of the temperature at the NV site. (b) Schematic of NV sensing of the local temperature and the Oersted field $\Delta B_{//}$ generated by the electric current flowing the VO$_2$ film. (c) An optically detected NV ESR spectrum measured when $B_{//}$ = 700 Oe, $I_{dc}$ = 0 mA at a base temperature of 295 K. (d) Electric current dependence of the IMT-induced local magnetic field $\Delta B_{//}$ at the NV site. The blue, green, and red curves correspond to the base temperatures of 320, 325, and 330 K, respectively. The jumps correspond to the resistive switching events as shown in Fig. 2(d). (e) Local temperature extracted from the NV ESR measurements as a function of the electric currents. The uncertainty on the measured temperature is ±1.2 K. The grey color stripe marks a temperature regime of 335 ±1 K, where the thermally-induced IMTs are expected to happen.

+1 and the m$_s$ = −1 spin states by an energy gap equal to $2\gamma B_{//}$, where $\gamma$ denotes the gyromagnetic ratio.[21] This three-level spin system can be optically read out through spin-dependent photoluminescence (PL) emitted by the NV center, where the $m_s = \pm 1$ spin states are more likely to be trapped by a non-radiative pathway through an intersystem crossing and back to the $m_s = 0$ ground state, yielding a significantly reduced PL intensity.[21] To perform the optically detected NV ESR measurements, an external magnetic field $B_{//}$ was applied and aligned along the NV-axis and the base temperature of the device was maintained at a constant value. A constant microwave current with a frequency $f$ was delivered by the Au stripline and a continuous green laser was focused on the NV site to initiate the NV spin to the $m_s = 0$ state. The spin-dependent PL generated by the NV center in the red wavelength range was measured using a single-photon detector.[37] The NV PL intensity was measured as a function of the microwave frequency $f$. When $f$ matches the NV ESR energies, the NV spin tends to flip from the $m_s = 0$ to the $m_s = \pm 1$ states, giving rise to two dips in the measured ESR spectrum. The Oersted field generated by the electric current flowing the VO$_2$ film can be obtained by the Zeeman splitting of the NV ESR frequencies:

$$\Delta B_{//} = \frac{\pi f_+ - \pi f_-}{\gamma} - B_{//} \quad (1)$$

where $f_\pm$ correspond to the NV ESR frequencies of the $m_s = 0 \leftrightarrow \pm 1$ transitions and $\Delta B_{//}$ is the component of the Oersted field that is parallel to the NV-axis as illustrated in Fig. 3(b). The local temperature $T_L$ at the NV site can be extracted from the NV ESR measurements as follows: $T_L = \frac{f_+ + f_-}{2b} - \frac{a}{b}$,[38–40] where $a$ and $b$ are fitting parameters equal to $2.8983 \pm 0.002$ GHz and $-88.9 \pm 5.8$ kHz/K, respectively. These values are obtained by measuring $f_\pm$ as a function of the sample temperature (see supplementary materials for details). Since the measurements were performed in a vacuum environment and due to the nanoscale proximity established between the NV center and the VO$_2$ sample, the NV center is well-thermalized with its immediate vicinity in the device (see supplementary materials for details) and the local temperature of the VO$_2$ sample can be measured. Figure 3(c) shows a typical NV ESR spectrum with $B_{//} = 700$ Oe and a base temperature of 295 K.

Figures 3(d) and 3(e) show the extracted Oersted field $\Delta B_{//}$ and the local temperature $T_L$ as a function of the applied electric current $I_{dc}$ at three different base temperatures of 320, 325, and 330 K. For low electric currents $I_{dc} < I_c$, VO$_2$ is in a homogeneous semi-insulating state and the electric current is sparsely distributed in the device, leading to a negligibly small Oersted field at the NV site. When the electric current reaches its critical value ($I_{dc} = I_c$), the IMT is electrically triggered, accompanied by the formation of a conducting filament in the VO$_2$ film. Since the electric current is then mainly concentrated in the conducting filament, the local current density and the Oersted field experienced by the NV center are significantly enhanced [Fig. 3(d)]. These sudden jumps in magnetic field correspond to switching events observed in the electrical transport measurements. The measured Oersted fields around the critical currents are in qualitative agreement with COMSOL simulations (see supplementary materials for details). In contrast to the jump-type variation of the Oersted field, the measured local temperature $T_L$ exhibits a gradual increase as a function of $I_{dc}$. It is worth mentioning that $T_L$ measured at different base temperatures reaches a similar value (~335 K) during resistive switching, demonstrating the thermal origin of the voltage-induced IMT observed in a pristine VO$_2$ film.[24,41] At a sufficiently large electric current, Joule heating along the current path locally increases the temperature of the VO$_2$ film to the critical value (~335 K) and triggers the formation of the conducting filaments.

In addition to Joule heating, it has been suggested that electric fields may also induce the IMT, but the origin of this effect is still debated.[25,42–48]. A number of reports have indicated that the IMT may be induced without reaching the IMT critical temperature.[17,18,41,49–52] However, due to the inhomogeneous nature of the IMT and the need for simulations of the current and temperature distribution in the sample, direct evidence for the non-thermally induced IMT remains elusive. Local measurements of temperature in the nanoscale during the IMT are of great value to confirm the possibility of non-thermal switching. Next, we applied the NV-based quantum sensing platform to ion-irradiated VO$_2$ thin film devices to access the mechanism of non-thermally induced IMT.[25] We employed a focused ion beam to irradiate gallium ions onto the VO$_2$ film in a ~2 μm wide region that connects the Au contacts (see supplementary materials for details). The gallium irradiation has interesting effects on the transport properties of the VO$_2$ thin film. First of all, the voltage-induced IMTs can be triggered at a much lower base temperature with reduced current/voltage/power as shown in Fig. 4(a). However, the resistance-temperature characteristics of the ion-irradiated VO$_2$ device remain largely the same, showing a similar $T_c$ as in the pristine sample [Fig. 4(b)]. Figure 4(c) shows the local temperature extracted from the NV ESR measurements as a function of $I_{dc}$. In stark contrast to the pristine VO$_2$ sample, the local temperature measured at the critical current barely changes from the base temperature which can be up to 35 K lower than $T_c$. It is important to note that the NV center is situated on the symmetry axis between the two contacts and is in contact with the conducting filament formed in the ion-irradiated VO$_2$ device (see supplementary materials for details). The absence of substantial heating both before and after the switching process demonstrates the non-thermal origin of the electrically-induced IMT in the ion-irradiated VO$_2$ sample.[25]

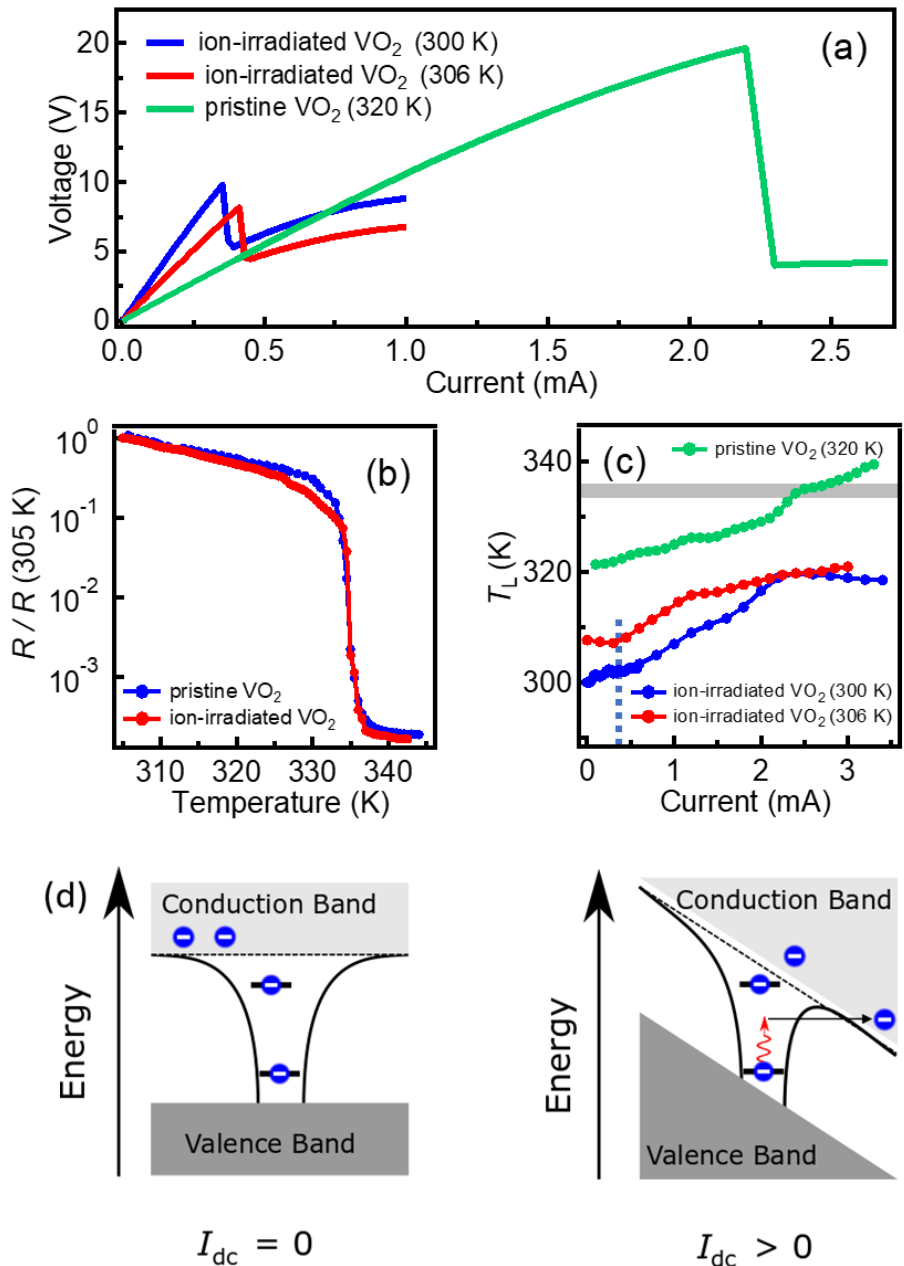

**Figure 4**. (a) Comparison of voltage-induced IMTs in pristine and ion-irradiated $VO_2$ films. (b) Variation of the resistance [$R/R$(305K)] of the pristine and ion-irradiated $VO_2$ devices measured as a function of the base temperature. The electrical resistance $R$ is normalized to the value measured at 305 K. (c) Local temperature extracted from the NV ESR measurements versus the applied electric current for pristine and ion-irradiated $VO_2$ devices. The uncertainty on the measured temperature is $\pm 1.2$ K. The grey color stripe marks a temperature regime of $335 \pm 1$ K, where the thermally-induced IMTs are expected to happen. The blue dash line indicates the critical current for the electrically-induced IMTs. (d) Schematic of the field-assisted carrier generation through a doping-driven IMT.

Our results are consistent with a previous study showing indirect evidence for non-thermal switching and support field-assisted carrier generation as the switching mechanism (see supplementary materials for details).[25,53,54] With a sufficiently large electric field, in-gap states created by ion beam irradiation could be electrically excited, emitting charge into the conduction band, as illustrated in Fig. 4(d). This increases the number of free carriers and causes the collapse of the Mott insulator state through a doping-driven IMT.[25] In comparison to thermally-induced resistive switching, we highlight that the critical currents and energy dissipations are significantly reduced in the doping-driven IMT, offering significant advantages for the development of energy-efficient neuromorphic circuits in a broad temperature range.

In summary, we have demonstrated NV centers as a sensitive local probe of the thermally- and non-thermally-induced IMTs in VO$_2$ devices. By measuring the local temperature and the magnetic field environment via the optically detected NV electron spin resonances, the underlying electrical phase transitions in proximal VO$_2$ devices could be accessed in a non-perturbative way at the nanoscale. This technique allowed us to obtain the first direct evidence for a non-thermally induced electrical IMT in a Mott insulator. The findings also have important implications for our understanding of the physics of Mott insulators and their applications. By employing patterned diamond nanostructures with shallowly implanted NV centers,[55–57] we expect that the local resolution of such NV quantum sensing platform could potentially reach the atomic scale, offering new opportunities to reveal the electrical and thermal behaviors in Mott insulators and many other quantum materials. The demonstrated coupling between NV centers and Mott insulators may also find applications in developing next-generation, hybrid neuromorphic devices.

**Acknowledgements**. The authors would like to thank Eric Lee-Wong for technical assistance and Paul Y. Wang for help with COMSOL simulations. The authors acknowledge Francesco Casola for fabricating the diamond nanobeams. N. J. M. and C. R. D. were supported by the Air Force Office of Scientific Research under award FA9550-20-1-0319 and the U. S. National Science Foundation under award ECCS-2029558. Y. K. and I. K. S. were supported by the Quantum Materials for Energy Efficient Neuromorphic Computing (Q-MEEN-C), an Energy Frontier

Research Center funded by the U.S. Department of Energy, Office of Science, Basic Energy Sciences under Award # DE-SC0019273.

# References

1. J. von Neumann, First draft of a report on the EDVAC, *IEEE Ann. Hist. Comput.* **15**, 27 (1993).
2. M. A. Nielsen and I. L. Chuang, *Quantum Computation and Quantum Information*. (Cambridge Univ. Press, Cambridge, 2000).
3. F. Arute *et al.*, Quantum supremacy using a programmable superconducting processor, *Nature* **574**, 505 (2019).
4. D. Monroe, Neuromorphic computing gets ready for the (really) big time, *Commun. ACM* **57**, 13 (2014).
5. J. del Valle *et al.*, Subthreshold firing in Mott nanodevices, *Nature* **569** 388 (2019).
6. H. T. Zhang *et al.*, Organismic materials for beyond von Neumann machines, *Appl. Phys. Rev* **7**, 011309 (2020).
7. J. Borghetti, G. S. Snider, P. J. Kuekes, J. J. Yang, D. R. Stewart, and R. S. Williams, Memristive switches enable stateful logic operations via material implication, *Nature* **464**, 873 (2010).
8. C. Mead, Neuromorphic electronic systems, *Proc. IEEE* **78**, 1629 (1990).
9. P. A. Merolla *et al.*, A million spiking-neuron integrated circuit with a scalable communication network and interface, *Science* **345**, 668 (2014).
10. S. H. Jo, T. Chang, I. Ebong, B. B. Bhadviya, P. Mazumder, and W. Lu, Nanoscale memristor device as synapse in neuromorphic systems, *Nano Lett.* **10**, 1297 (2010).
11. M. Imada, A Fujimori, and Y. Tokura, Metal-insulator transitions, *Rev. Mod. Phys.* **70**, 1039 (1998).
12. J. H. Park, J, M. Coy, T. S, Kasirga, C, Huang, Z, Fei, S, Hunter, and D. H. Cobden, Measurement of a solid-state triple point at the metal-insulator transition in $VO_2$, *Nature* **500**, 431 (2013).
13. P. Stoliar, J. Tranchant, B. Corraze, E. Janod, M.-P. Besland, F. Tesler, M. Rozenberg, and L. Cario, A Leaky-integrate-and-fire neuron analog realized with a Mott insulator, *Adv. Funct. Mater.* **27**, 1604740 (2017).
14. J. H. Ngai, F. J. Walker, and C. H. Ahn, Correlated oxide physics and electronics, *Annu. Rev. Mater. Res.* **44**, 1 (2014).
15. T. M. Dao, P. S. Mondal, Y. Takamura, E. Arenholz, and J. Lee, Metal-insulator transition in low dimensional $La_{0.75}Sr_{0.25}VO_3$ thin films, *Appl. Phys. Lett.* **99**, 112111 (2011).
16. J. S. Brockman, L. Gao, B. Hughes, C. T. Rettner, M. G. Samant, K. P. Roche, and S. S. P. Parkin, Subnanosecond incubation times for electric-field induced metallization of a correlated electron oxide, *Nat. Nanotechnol.* **9**, 453 (2014).
17. G. Stefanovich, A. Pergament, and D. Stefanovich, Electrical switching and Mott transition in $VO_2$, *J. Phys. Condens. Matter.* **12**, 8837 (2000).
18. P. Diener, E. Janod, B. Corraze, M. Querré, C. Adda, M. Guilloux-Viry, S. Cordier, A. Camjayi, M. Rozenberg, M. P. Besland, and L. Cario, How a dc electric field drives Mott insulators out of equilibrium, *Phys. Rev. Lett.* **121**, 16601 (2018).
19. D. Li, A. A. Sharma, D. K. Gala, N. Shukla, H. Paik, S. Datta, D. G. Schlom, J. A. Bain, and M. Skowronski, Joule heating-induced metal–insulator transition in epitaxial $VO_2/TiO_2$ devices, *ACS Appl. Mater. Interfaces.* **8**, 12908 (2016).
20. F. Giorgianni, J. Sakai, and S. Lupi, Overcoming the thermal regime for the electric-field driven Mott transition in vanadium sesquioxide, *Nat. Commun.* **10**, 1159 (2019).
21. L. Rondin, J.-P. Tetienne, T Hingant, J.-F. Roch, P Maletinsky, and V Jacques, Magnetometry with nitrogen-vacancy defects in diamond, *Reports Prog. Phys.* **77**, 056503 (2014).

22. D. Lee *et al.*, Isostructural metal-insulator transition in $VO_2$, *Science* **362**, 1037 (2018).
23. A. Mansingh and R. Singht, The mechanism of electrical threshold switching in $VO_2$, *crystals. J. Phys. C: Solid St. Phys* **13**, 5725 (1980).
24. A. Zimmers, L. Aigouy, M. Mortier, A. Sharoni, S. Wang, K. G. West, J. G. Ramirez, and I. K. Schuller, Role of thermal heating on the voltage induced insulator-metal transition in $VO_2$, *Phys. Rev. Lett.* **110**, 056601 (2013).
25. Y. Kalcheim, A. Camjayi, J. del Valle, P. Salev, M. Rozenberg, and I. K. Schuller, Non-thermal resistive switching in Mott insulators, *Nat. Commun.* **11**, 2985 (2020).
26. M. J. Burek, N. P. de Leon, B. J. Shields, B. J. M. Hausmann, Y. Chu, Q. Quan, A. S. Zibrov, H. Park, M. D. Lukin, and M. Lončar, Free-standing mechanical and photonic nanostructures in single-crystal diamond, *Nano Lett.* **12**, 6084 (2012).
27. C. Du. *et al.*, Control and local measurement of the spin chemical potential in a magnetic insulator, *Science* **357**, 195 (2017).
28. G. D. Fuchs, V. V. Dobrovitski, D. M. Toyli, F. J. Heremans, and D. D. Awschalom, Gigahertz dynamics of a strongly-driven single quantum spin, *Science* **326**, 1520 (2009).
29. J. Del Valle, Y. Kalcheim, J. Trastoy, A. Charnukha, D. N. Basov, and I. K. Schuller, Electrically induced multiple metal-insulator transitions in oxide nanodevices, *Phys. Rev. Appl.* **8**, 054041 (2017).
30. M. S. Grinolds, M. Warner, K. De Greve, Y. Dovzhenko, L. Thiel, R. L. Walsworth, S. Hong, P. Maletinsky, and A. Yacoby, Subnanometre resolution in three-dimensional magnetic resonance imaging of individual dark spins, *Nat. Nanotechnol.* **9**, 279 (2014).
31. G. Balasubramanian *et al.*, Nanoscale imaging magnetometry with diamond spins under ambient conditions, *Nature* **455**, 648 (2008).
32. G. Kucsko, P. C. Maurer, N. Y. Yao, M. Kubo, H. J. Noh, P. K. Lo, H. Park, and M. D. Lukin, Nanometre-scale thermometry in a living cell, *Nature* **500**, 54 (2013).
33. D. M. Toyli, C. F. de las Casas, D. J. Christle, V. V. Dobrovitski, and D. D. Awschalom, Fluorescence thermometry enhanced by the quantum coherence of single spins in diamond, *Proc. Natl. Acad. Sci.* **110**, 8417 (2013).
34. F. Casola, T. van der Sar, and A. Yacoby, Probing condensed matter physics with magnetometry based on nitrogen-vacancy centres in diamond, *Nat. Rev. Mater.* **3**, 17088 (2018).
35. A. Laraoui *et al.*, Imaging thermal conductivity with nanoscale resolution using a scanning spin probe, *Nat. Commun.* **6**, 8954 (2015).
36. M. M. Qazilbash *et al.*, Mott transition in $VO_2$ revealed by infrared spectroscopy and nano-imaging, *Science* **318**, 1750 (2007).
37. E. Lee-Wong, R. L. Xue, F. Y. Ye, A. Kreisel, T. van der Sar, A. Yacoby, and C. H. R. Du, Nanoscale detection of magnon excitations with variable wavevectors through a quantum spin sensor, *Nano Lett.* **20**, 3284 (2020).
38. P. Neumann *et al.*, High-precision nanoscale temperature sensing using single defects in diamond, *Nano Lett.* **13**, 2738 (2013).
39. V. M. Acosta, E. Bauch, M. P. Ledbetter, A. Waxman, L. S. Bouchard, and D. Budker, Temperature dependence of the nitrogen-vacancy magnetic resonance in diamond, *Phys. Rev. Lett.* **104**, 070801 (2010).
40. N. Wang *et al.*, Magnetic criticality enhanced hybrid nanodiamond thermometer under ambient conditions, *Phys. Rev. X* **8**, 011042 (2018).

41. I. Valmianski, P. Y. Wang, S. Wang, J. Gabriel Ramirez, S. Guénon, and I. K. Schuller, Origin of the current-driven breakdown in vanadium oxides: thermal versus electronic, *Phys. Rev. B* **98**, 195144 (2018).
42. T. Oka, R. Arita, and H. Aoki, Breakdown of a Mott insulator: a nonadiabatic tunneling mechanism, *Phys. Rev. Lett.* **91**, 66406 (2003).
43. H. Yamakawa *et al.*, Mott transition by an impulsive dielectric breakdown, *Nat. Mater.* **16**, 1100 (2017).
44. B. Mayer *et al.*, Tunneling breakdown of a strongly correlated insulating state in $VO_2$ by intense multiterahertz excitation, *Phys. Rev. B* **91**, 235113 (2015).
45. T. Oka and H. Aoki, Ground-state decay rate for the Zener breakdown in band and Mott insulators, *Phys. Rev. Lett.* **95**, 137601 (2005).
46. B. Wu *et al.*, Electric-field-driven phase transition in vanadium dioxide, *Phys. Rev. B* **84**, 241410 (2011).
47. W.-R. Lee and K. Park, Dielectric breakdown via emergent nonequilibrium steady states of the electric-field-driven Mott insulator, *Phys. Rev. B* **89**, 205126 (2014).
48. N. Sugimoto, S. Onoda, and N. Nagaosa, Field-induced metal-insulator transition and switching phenomenon in correlated insulators, *Phys. Rev. B* **78**, 155104 (2008).
49. P. Stoliar, M. Rozenberg, E. Janod, B. Corraze, J. Tranchant, and L. Cario, Nonthermal and purely electronic resistive switching in a Mott memory, *Phys. Rev. B.* **90**, 45146 (2014).
50. E. Janod *et al.*, Resistive switching in Mott insulators and correlated systems, *Adv. Funct. Mater.* **25**, 6287 (2015).
51. P. Stoliar, L. Cario, E. Janod, B. Corraze, C. Guillot-Deudon, S. Salmon-Bourmand, V. Guiot, J. Tranchant, and M. Rozenberg, Universal electric-field driven resistive transition in narrow-gap Mott insulators, *Adv. Mater.* **25**, 3222 (2013).
52. G. Gopalakrishnan, D. Ruzmetov, and S. Ramanathan, On the triggering mechanism for the metal–insulator transition in thin film $VO_2$ devices: electric field versus thermal effects, *J. Mater. Sci.* **44**, 5345 (2009).
53. S. D. Ganichev, E. Ziemann, W. Prettl, I. N. Yassievich, A. A. Istratov, and E. R. Weber, Distinction between the Poole-Frenkel and tunneling models of electric-field-stimulated carrier emission from deep levels in semiconductors, *Phys. Rev. B* **61**, 10361 (2020).
54. J. Frenkel, On pre-breakdown phenomena in insulators and electronic semi-conductors, *Phys. Rev.* **54**, 647 (1938).
55. L. Thiel, D. Rohner, M. Ganzhorn, P. Appel, E. Neu, B. Müller, R. Kleiner, D. Koelle, and P. Maletinsky, Quantitative nanoscale vortex imaging using a cryogenic quantum magnetometer, *Nat. Nanotechnol.* **11**, 677 (2016).
56. M. Pelliccione, Alec Jenkins, P. Ovartchaiyapong, C. Reetz, E. Emmanouilidou, N. Ni, and A. C. Bleszynski Jayich, Scanned probe imaging of nanoscale magnetism at cryogenic temperatures with a single-spin quantum sensor, *Nat. Nanotechnol.* **11**, 700 (2016).
57. S. J. Devience *et al.*, Nanoscale NMR spectroscopy and imaging of multiple nuclear species, *Nat. Nanotechnol.* **10**, 129 (2015).